\documentclass[sigconf,nonacm]{acmart}

\usepackage{algorithmic}
\usepackage{graphicx}
\usepackage{textcomp}
\usepackage{xcolor}
\usepackage{comment}
\usepackage{enumitem}

\usepackage{tikz}
\usetikzlibrary{positioning,arrows.meta,shapes.geometric,calc}
\usepackage{dblfloatfix}
\usepackage{tabularx}

\usepackage{booktabs}
\usepackage[tableposition=top, skip=5pt]{caption}
\usepackage{multicol}
\usepackage{multirow}
\usepackage{array}
\usepackage{ragged2e}
\newcolumntype{L}[1]{>{\RaggedRight\let\newline\\\arraybackslash\hspace*{0pt}}p{#1}}
\newcolumntype{C}[1]{>{\centering\let\newline\\\arraybackslash\hspace*{0pt}}p{#1}}
\newcolumntype{R}[1]{>{\RaggedLeft\let\newline\\\arraybackslash\hspace*{0pt}}p{#1}}

\usepackage{pgfplots}
\pgfplotsset{compat=1.18}
\usepackage{float}
\usepackage{url}

\begin{document}

\title{GenAI in Software Engineering: The Role of Technology Acceptance Models}

\author{Oscar Johansson}
\affiliation{%
  \institution{Blekinge Institute of Technology}
  \city{Karlskrona}
  \country{Sweden}
}
\email{oscar.johansson@bth.se}

\author{Jürgen Börstler}
\affiliation{%
  \institution{Blekinge Institute of Technology}
  \city{Karlskrona}
  \country{Sweden}
}
\email{jurgen.borstler@bth.se}

\author{Nauman bin Ali}
\affiliation{%
  \institution{Blekinge Institute of Technology}
  \city{Karlskrona}
  \country{Sweden}
}
\email{nauman.ali@bth.se}

\renewcommand{\shortauthors}{Johansson et al.}

\begin{abstract}

\noindent \textbf{Context:} Many organizations are keen to incorporate generative~AI (GenAI) into their software development processes. Technology acceptance models, such as the Unified Theory of Acceptance and Use of Technology (UTAUT), are traditionally used to identify individual-level barriers to the acceptance of new technologies and can facilitate the transition to GenAI. However, UTAUT has seen limited use within software engineering (SE) research.

\noindent \textbf{Objective:} Using UTAUT as an example, to identify key areas for future research on GenAI acceptance, including the role of Bayesian approaches for data analysis.

\noindent \textbf{Method:} We review foundational and SE-specific literature on UTAUT and analyze its emerging applications for GenAI in SE.

\noindent \textbf{Results:} We identify three priorities for future research:
(1) identifying and refining constructs to account for GenAI's nature and transformational impact;
(2) improving operationalization practices to strengthen construct validity and cross-study comparability; and
(3) incorporating Bayesian analysis to support small-sample inference by integrating prior knowledge, iterative model updating, and simulation of scenarios.

\noindent \textbf{Conclusion:} UTAUT is a suitable candidate to combine with Bayesian analysis for practical insights on individual-level barriers to GenAI use in SE, but additional theories should be considered.

\end{abstract}

\keywords{Acceptance Behavior, Bayesian Analysis, Theory Use in Software Engineering, UTAUT}

\maketitle

\section{Introduction}

Generative artificial intelligence (GenAI) is transforming software engineering (SE) practices \cite{nguyen-ducGenerativeArtificialIntelligence2025a, strayWhatGenerativeAI2025a}. Organizations are increasingly seeking to use these technologies to improve their software development process \cite{ebert2023generative,sauvola2024future}. Furthermore, GenAI is becoming more than just a tool automating recurring manual tasks. It is increasingly being perceived as a human-like collaborator rather than just an inanimate tool \cite{zakharovAISoftwareEngineering2025}. This shift is largely due to recent advances in GenAI, which allow us to interact with these models in ways that resemble human communication. As GenAI advances, its impact cannot be isolated to a limited scope of tool adoption. It will continue to redefine the roles of software engineers and the practice of software engineering \cite{nguyen-ducGenerativeArtificialIntelligence2025a, strayWhatGenerativeAI2025a}. Amidst this disruptive transformation, understanding why and how software engineers accept and engage with this technology is crucial for successful GenAI adoption and achieving business competitiveness.

Technology acceptance models can play a crucial role in identifying barriers to the acceptance and use of GenAI among software engineers. Both the Technology Acceptance Model (TAM) \cite{davisPerceivedUsefulnessPerceived1989} and the Unified Theory of Acceptance and Use of Technology (UTAUT)~\cite{venkateshUserAcceptanceInformation2003} are extensively used in other disciplines\footnote{According to a Scopus search on April 29, 2026, the four most cited publications on TAM have 76570 citations by 57525 unique documents. The two most cited publications on UTAUT have 45857 citations by 38192 unique documents.}, with limited use within SE \cite{borstlerAcceptanceBehaviorTheories2024}.
We find that UTAUT is especially suitable for the SE context, as it extends TAM by focusing on organizational contexts and introducing key constructs to enhance the model's explanatory power. Moreover, a recent meta-analysis based on 1451 empirical articles that employ UTAUT \cite{blutMetaAnalysisUnifiedTheory2022} provides strong evidence for the relevance of UTAUT constructs in explaining technology acceptance and use. Despite its advantages, UTAUT has seen minimal use within SE research compared to TAM \cite{borstlerAcceptanceBehaviorTheories2024}. Only recently has this begun to change, as there is an increase in studies using UTAUT to specifically study GenAI acceptance in SE \cite{russoNavigatingComplexityGenerative2024, lambiaseInvestigatingRoleCultural2025, shaoEmpiricalAnalysisGenerative2026c, elinzanoChatGPTAcceptanceUse2025}. In line with the traditional UTAUT approach \cite{venkateshUserAcceptanceInformation2003, venkateshConsumerAcceptanceUse2012}, these studies examine the factors influencing GenAI acceptance among software engineers using a Likert-scale quantitative instrument for data collection and frequentist statistics for analysis. 

However, the traditional UTAUT approach has certain limitations when applied in empirical software engineering. For instance, it can be challenging to obtain a sufficiently large sample for data analysis \cite{ghaziSurveyResearchSoftware2019}, which limits widespread application. Additionally, frequentist analysis using statistical hypothesis testing may oversimplify analysis, and interpreting the practical significance of effect sizes and confidence intervals can be challenging \cite{furiaBayesianDataAnalysis2019}. Thus, we want to look beyond the traditional approach as we explore the suitability of UTAUT for studying GenAI acceptance in SE.

In this paper, we argue for a Bayesian approach grounded in UTAUT to study GenAI acceptance in SE. The Bayesian approach can address the same questions as the frequentist approach in empirical SE, while overcoming many of its limitations and offering additional insights \cite{furiaBayesianDataAnalysis2019}. We make three primary contributions. First, we conduct a literature review and identify an initial set of ten relevant constructs for GenAI acceptance in SE grounded in UTAUT. Second, we identify three priorities for future research based on our review findings. Third, we outline how a Bayesian approach can provide software engineers with practical insights to facilitate the transition to GenAI.

The remainder of the paper is structured as follows. In Section~\ref{sec:UTAUT_AI}, we review a subset of the UTAUT literature to explore which constructs may explain GenAI acceptance in SE. The review goes beyond the original UTAUT \cite{venkateshUserAcceptanceInformation2003} and UTAUT2 \cite{venkateshConsumerAcceptanceUse2012} to get a more complete understanding of our domain and how the theory has evolved over time. In Section \ref{sec:Bayesian}, we introduce the Bayesian approach and its benefits compared to frequentist analysis. In Section \ref{sec:Example}, we present a fictitious example to illustrate how software engineers can apply a Bayesian approach grounded in UTAUT to identify barriers to GenAI use and evaluate different interventions to address them.

\section{UTAUT for GenAI Acceptance in SE} \label{sec:UTAUT_AI}

\begin{table*}[h!]
    \centering
    \small
    \caption{Predictors of the constructs Behavioral Intention and Technology Use together with their definitions and example question items from the literature.}
    \renewcommand{\arraystretch}{1.25}
    \resizebox{0.99\textwidth}{!}{%
    \begin{tabular}{L{1.7cm} p{4.5cm} C{0.9cm} C{0.9cm} C{0.9cm} C{0.9cm} L{4.65cm}}
        \toprule
        \textbf{Construct} & \textbf{Definition} & \textbf{UTAUT \cite{venkateshUserAcceptanceInformation2003}} & \textbf{UTAUT2 \cite{venkateshConsumerAcceptanceUse2012}} & \textbf{Blut et al. \cite{blutMetaAnalysisUnifiedTheory2022}} & \textbf{Our} & \textbf{Example Measurement Items} \\
        \midrule 
        Performance Expectancy & ``[T]he degree to which an individual believes that using the system will help him or her to attain gains in job performance.'' \cite{venkateshUserAcceptanceInformation2003} & X & X & X & X &
        {\vspace{-1.1\baselineskip} \begin{itemize}[label={-}, nosep, leftmargin=*]
        \item ``Using [GenAI] to support my job increases my productivity'' \cite{lambiaseInvestigatingRoleCultural2025}
        \item ``[GenAI] improve the quality of my software development work'' \cite{shaoEmpiricalAnalysisGenerative2026c}
        \end{itemize}}
        \\ 
        Effort Expectancy & ``[T]he degree of ease associated with the use of the system.'' \cite{venkateshUserAcceptanceInformation2003} & X & X & X & X & {\vspace{-1.1\baselineskip} \begin{itemize}[label={-}, nosep, leftmargin=*]
        \item ``I find [GenAI] easy to use in supporting my job'' \cite{lambiaseInvestigatingRoleCultural2025}
        \item ``It is straightforward to use [GenAI] for software tasks'' \cite{shaoEmpiricalAnalysisGenerative2026c}
        \end{itemize}} \\
        Social Influence &  ``[T]he degree to which an individual perceives that important others believe he or she should use the new system.''~\cite{venkateshUserAcceptanceInformation2003} & X & X & X & X & {\vspace{-1.1\baselineskip} \begin{itemize}[label={-}, nosep, leftmargin=*]
        \item ``People who are important to me think that I should use [GenAI]'' \cite{russoNavigatingComplexityGenerative2024}
        \item ``My organisation/institution recognises proficiency in [GenAI] as an important skill'' \cite{shaoEmpiricalAnalysisGenerative2026c}
        \end{itemize}} \\
        Facilitating Conditions & ``[T]he degree to which an individual believes that an organizational and technical infrastructure exists to support use of the system.'' \cite{venkateshUserAcceptanceInformation2003} & X & X & X & X & {\vspace{-1.1\baselineskip} \begin{itemize}[label={-}, nosep, leftmargin=*]
        \item ``I have the knowledge necessary to use [GenAI] to support my job'' \cite{lambiaseInvestigatingRoleCultural2025}
        \item ``My organization offers sufficient training sessions to enhance our skills in using [GenAI]'' \cite{russoNavigatingComplexityGenerative2024}
        \end{itemize}} \\
        Price Value & ``[T]he individual’s cognitive tradeoff between the perceived benefits of the applications and the monetary cost for using them'' \cite{venkateshConsumerAcceptanceUse2012} & -- & (X)* & X & -- & {\vspace{-1.1\baselineskip} \begin{itemize}[label={-}, nosep, leftmargin=*]
        \item ``[GenAI] that I use to support my job are reasonably priced'' \cite{lambiaseInvestigatingRoleCultural2025}
        \item ``[GenAI] that I use to support my job are a good value for the money'' \cite{lambiaseInvestigatingRoleCultural2025}
        \end{itemize}} \\
        Hedonic Motivation & ``[T]he fun or pleasure derived from using a technology'' \cite{venkateshConsumerAcceptanceUse2012} & -- & X & X & X & {\vspace{-1.1\baselineskip} \begin{itemize}[label={-}, nosep, leftmargin=*]
        \item ``Using [GenAI] to support my job is fun'' \cite{lambiaseInvestigatingRoleCultural2025}
        \item ``Using [GenAI] stimulates my curiosity'' \cite{kimNotMerelyUseful2025}
        \end{itemize}} \\
        Habit & ``[T]he extent to which people tend to perform behavior automatically because of learning'' \cite{venkateshConsumerAcceptanceUse2012} & -- & X & X & X & {\vspace{-1.1\baselineskip} \begin{itemize}[label={-}, nosep, leftmargin=*]
        \item ``Using [GenAI] to support my job has become natural to me'' \cite{lambiaseInvestigatingRoleCultural2025}
        \item ``The use of [GenAI] to support my job has become a habit for me'' \cite{lambiaseInvestigatingRoleCultural2025}
        \end{itemize}} \\
        Technology Compatibility & ``The degree to which an innovation is perceived as being consistent with existing values, needs, and experiences of potential adopters.'' \cite{blutMetaAnalysisUnifiedTheory2022} & -- & -- & X & X & {\vspace{-1.1\baselineskip} \begin{itemize}[label={-}, nosep, leftmargin=*]
        \item ``I think that using [GenAI] fits well with the way I like to work'' \cite{russoNavigatingComplexityGenerative2024}
        \item ``Using [GenAI] is compatible with my current situation'' \cite{kimNotMerelyUseful2025}
        \end{itemize}} \\
        User Education & ``The education level of the user.'' \cite{blutMetaAnalysisUnifiedTheory2022} & -- & -- & X & -- & \textit{None found} \\
        Personal Innovativeness & ``[A]n individual characteristic reflecting a willingness to try out any new technology.'' \cite{blutMetaAnalysisUnifiedTheory2022} & -- & -- & X & X & {\vspace{-1.1\baselineskip} \begin{itemize}[label={-}, nosep, leftmargin=*]
        \item ``I like to experiment with new technologies.'' \cite{russoNavigatingComplexityGenerative2024}
        \item ``Among my peers, I am usually the first to try out new technologies like [GenAI]'' \cite{russoNavigatingComplexityGenerative2024}
        \end{itemize}} \\
        Costs of Technology & ``The extent to which a user perceives that using a technology is costly.'' \cite{blutMetaAnalysisUnifiedTheory2022} & -- & -- & X & X & {\vspace{-1.1\baselineskip} \begin{itemize}[label={-}, nosep, leftmargin=*]
        \item ``I think that [using GenAI] in programming tasks [is] expensive'' \cite{parkContinuanceUseAI2025}
        \item ``There are financial barriers to using [GenAI] in programming tasks'' \cite{parkContinuanceUseAI2025}
        \end{itemize}} \\
        Trust & ``[T]he attitude that an agent will help achieve an individual’s goals in a situation characterized by uncertainty and vulnerability'' \cite{choudhuriWhatGuidesOur2025b} & -- & -- & -- & X & {\vspace{-1.1\baselineskip} \begin{itemize}[label={-}, nosep, leftmargin=*]
        \item ``I feel safe that when I rely on the [GenAI] I will get the right answers'' \cite{hoffmanMeasuresExplainableAI2023} as cited by \cite{choudhuriWhatGuidesOur2025b}
        \item ``I am confident in the [GenAI]. I feel that it works well'' \cite{hoffmanMeasuresExplainableAI2023} as cited by~\cite{choudhuriWhatGuidesOur2025b}
        \end{itemize}} \\
        \bottomrule
        \multicolumn{7}{l}{\footnotesize{*Introduced specifically for consumer technology setting. Venkatesh et al.~\cite{venkateshUnifiedTheoryAcceptance2016} explicitly suggest omitting price value in organizational settings.}}
    \end{tabular}}
    \label{tab:constructs}
\end{table*}

UTAUT introduces several constructs to explain technology acceptance and use. By measuring these constructs, it is possible to predict the extent to which individuals are likely to accept a given technology, as well as potential barriers to its adoption. The theory originates from the domain of information systems, and the original UTAUT \cite{venkateshUserAcceptanceInformation2003} comprises four constructs: performance expectancy, effort expectancy, social influence, and facilitating conditions. Later, the theory was extended to consumers in UTAUT2 \cite{venkateshConsumerAcceptanceUse2012}, by incorporating three additional constructs: price value, hedonic motivation, and habit. Recently, Blut et al.~\cite{blutMetaAnalysisUnifiedTheory2022} conducted a meta-analysis on the usage of the UTAUT and UTAUT2 models. Their research revealed a wide variety of adaptations and enhancements made to the original UTAUT/UTAUT2. In total, they identified 72 constructs derived from 1,451 papers. Based on their findings, they recommend including four additional constructs: technology compatibility, user education, personal innovativeness (also known as resistance to change), and costs of technology. This amounts to eleven constructs. 

Initial studies on GenAI acceptance in SE also focus on these constructs. We retrieved seven studies \cite{choudhuriWhatGuidesOur2025b, elinzanoChatGPTAcceptanceUse2025, kimNotMerelyUseful2025, lambiaseInvestigatingRoleCultural2025, parkContinuanceUseAI2025, russoNavigatingComplexityGenerative2024, shaoEmpiricalAnalysisGenerative2026c} on GenAI acceptance in SE from a snowball search on UTAUT \cite{venkateshUserAcceptanceInformation2003}, UTAUT2 \cite{venkateshConsumerAcceptanceUse2012}, and TAM \cite{davisPerceivedUsefulnessPerceived1989}. The search was conducted in Scopus on February 13, 2026, limited to papers in the computer science subject area published after 2022. The studies investigate the predictive power of UTAUT constructs on behavioral intention and technology use. Statistical significance has been shown for performance expectancy \cite{lambiaseInvestigatingRoleCultural2025, elinzanoChatGPTAcceptanceUse2025, parkContinuanceUseAI2025}, social influence \cite{shaoEmpiricalAnalysisGenerative2026c, elinzanoChatGPTAcceptanceUse2025}, facilitating conditions \cite{lambiaseInvestigatingRoleCultural2025}, hedonic motivation \cite{kimNotMerelyUseful2025}, habit \cite{lambiaseInvestigatingRoleCultural2025}, and technology compatibility \cite{russoNavigatingComplexityGenerative2024}. However, the results of the studies are not fully consistent with one another. Moreover, statistical significance has not been shown for effort expectancy \cite{lambiaseInvestigatingRoleCultural2025, shaoEmpiricalAnalysisGenerative2026c, elinzanoChatGPTAcceptanceUse2025}, personal innovativeness \cite{russoNavigatingComplexityGenerative2024}, and costs of technology \cite{parkContinuanceUseAI2025}. Price value and user education have not been analyzed in any of the studies\footnote{Lambiase et al.~\cite{lambiaseInvestigatingRoleCultural2025} include price value in their instrument, but omit it from the analysis.}.

In addition to general constructs, UTAUT may also be extended with AI-specific constructs. Both Blut et al.~\cite{blutMetaAnalysisUnifiedTheory2022} and Venkatesh \cite{venkateshAdoptionUseAI2022} address the potential need for AI-specific constructs due to issues such as black-box models, model errors, model bias, and greater trust in human judgment \cite{venkateshAdoptionUseAI2022}. Recent SE research fills a part of this gap by introducing the AI-specific construct trust \cite{choudhuriWhatGuidesOur2025b, elinzanoChatGPTAcceptanceUse2025}, where the output quality of the AI-based tool has a large influence on trust \cite{choudhuriWhatGuidesOur2025b}. Both studies show that the predictive power of trust on behavioral intention is statistically significant. However, there may be additional constructs that are worth exploring from related theories. 

Among the seven studies from the snowball search, we did not find a clear consensus on which constructs to measure. Thus, we establish our first research priority:

\vspace{0.1\baselineskip}
\noindent
\fbox{\begin{minipage}[!]{0.98\columnwidth}%
\textbf{Priority 1:} To identify and refine the most meaningful constructs for studying GenAI acceptance in SE.
\end{minipage}}
\vspace{0.1\baselineskip}

We can use TAM's Seven Pillars Framework \cite{davisTechnologyAcceptanceModel2024} to identify relevant constructs and build an initial model. The framework provides general guidance for how to customize a technology acceptance model for a specific context. Among other principles, the framework highlights that constructs from the original theory (e.g., TAM or UTAUT) should be retained to enable comparison between empirical studies, that meta-analysis studies should be considered when prioritizing constructs, and that there should be a balance between the simplicity and richness of the new model. Following these principles, we emphasize ten of the twelve constructs introduced in this section, excluding price value and user education. The constructs are summarized in Table~\ref{tab:constructs}, with definitions, sources, and example measurement items. Price value is omitted, as suggested by Venkatesh et al.~\cite{venkateshUnifiedTheoryAcceptance2016} and Lambiase et al.~\cite{lambiaseInvestigatingRoleCultural2025}, since it only applies to settings where the users of the technology pay for its use themselves. Instead, we include costs of technology, as suggested by Blut et al.~\cite{blutMetaAnalysisUnifiedTheory2022}. In organizational settings, users might still have relevant opinions on how their organizations invest in technology. User education is omitted since Blut et al.'s meta-analysis indicates a small effect in organizational settings \cite{blutMetaAnalysisUnifiedTheory2022}. 

The previous argumentation illustrates a general application of the framework. The next step is to further explore constructs specific to GenAI acceptance in SE. Both by drawing on existing literature and applying UTAUT in practice.

Moreover, a suitable set of constructs is not sufficient by itself. The operationalization of the constructs also influences the suitability of UTAUT. Traditionally, Likert-scale instruments are used for data collection \cite{venkateshUserAcceptanceInformation2003, venkateshConsumerAcceptanceUse2012}. This is also the case in the SE-specific studies we have reviewed. 
We have listed example measurement items from those instruments in Table~\ref{tab:constructs}. In terms of operationalization, there are cases where the instrument remains close to the original UTAUT instrument \cite{lambiaseInvestigatingRoleCultural2025}, where measurement items have been refined for the SE context \cite{parkContinuanceUseAI2025}, and where the instrument is combined with focus groups for more nuanced insights \cite{elinzanoChatGPTAcceptanceUse2025}. Based on this observation, we establish a second research priority: 

\vspace{0.1\baselineskip}
\noindent
\fbox{\begin{minipage}[!]{0.98\columnwidth}%
\textbf{Priority 2:} To identify the most suitable way to operationalize constructs for the specific context of GenAI acceptance in SE.
\end{minipage}}
\vspace{0.1\baselineskip}

As UTAUT has seen limited use in SE contexts \cite{borstlerAcceptanceBehaviorTheories2024}, it is important to revisit the definitions and operationalizations of the constructs and make them meaningful in a modern SE context. This needs to be done carefully, though, so that users of the theory can still draw on the theoretical foundation of UTAUT and the wealth of empirical research conducted in the past. Currently, each study we have reviewed on GenAI acceptance in SE uses its own instrument. As a next step, we suggest developing a more unified instrument to facilitate cross-study comparisons and strengthen construct validity over time. 

The approach to data analysis may also need to evolve. GenAI changes constantly with rapid improvements and new functionality. Additionally, software engineering teams can be small and highly specialized. Our findings indicate that a frequentist approach to data analysis, such as PLS-SEM, is the most commonly used method for GenAI acceptance in SE. However, this approach requires redoing the analysis whenever new data becomes available, and it demands a sufficiently large sample size, which may not always be feasible in our context \cite{ghaziSurveyResearchSoftware2019}. Therefore, we establish a third research priority: 

\vspace{0.1\baselineskip}
\noindent
\fbox{\begin{minipage}[!]{0.98\columnwidth}%
\textbf{Priority 3:} To explore Bayesian approaches to data analysis for more flexibility and practical insights.
\end{minipage}}
\vspace{0.1\baselineskip}

As we did not find any studies on GenAI acceptance in SE that use Bayesian approaches, we will further discuss the potential of Bayesian approaches as an alternative to frequentist analysis in the next section.

\section{A Bayesian Approach Grounded in UTAUT}
\label{sec:Bayesian}

Studying GenAI acceptance involves not only determining what data to collect but also how to analyze that data. Although the frequentist approach remains the common choice for studying GenAI acceptance in SE \cite{choudhuriWhatGuidesOur2025b, elinzanoChatGPTAcceptanceUse2025, kimNotMerelyUseful2025, lambiaseInvestigatingRoleCultural2025, parkContinuanceUseAI2025, russoNavigatingComplexityGenerative2024, shaoEmpiricalAnalysisGenerative2026c}, a Bayesian approach offers several advantages \cite{furiaBayesianDataAnalysis2019}. In this section, we advocate for considering Bayesian analysis when operationalizing UTAUT. Bayesian approaches have been explored within other domains \cite{kimUnderstandingFactorsInfluencing2025a, lafifaRestructuringExpandingTechnology2012, salarzadehjenatabadiTestingStudentsElearning2017, kambleMachineLearningBased2021}, but are not commonly used for technology acceptance within SE. Thus, we will provide a brief introduction to the Bayesian approach, outline its benefits for studying GenAI acceptance in SE, and discuss the role of UTAUT.

Bayesian analysis is a statistical method in which the probability of a hypothesis (i.e., \textit{belief}) is updated as new data (i.e., \textit{evidence}) is observed. The approach has three main parts: a prior, a likelihood, and a posterior. Given a belief $B$ and evidence $E$. The prior is the initial assumption about the belief $P(B)$. The likelihood is the probability of the evidence given the belief $P(E|B)$. The posterior is then the updated probability distribution for the belief given the evidence $P(B|E)$, and is calculated using Bayes' theorem, that is,
\begin{equation*}
    P(B|E) = \frac{P(E|B) \cdot P(B)}{P(E)},
\end{equation*}
with $P(E)$ as a normalizing constant. Since calculating $P(B|E)$ directly can be computationally expensive, the posterior distribution is often estimated numerically using Markov chain Monte Carlo (MCMC) instead \cite{mcelreathStatisticalRethinkingBayesian2020}. While this is only a brief introduction to Bayesian analysis, a more detailed introduction would be outside the scope of this paper. We refer to McElreath \cite{mcelreathStatisticalRethinkingBayesian2020} or Gelman~et~al.~\cite{gelmanBayesianDataAnalysis2013} for a more comprehensive introduction. 

Within technology acceptance research, there are two distinct research directions using Bayesian approaches. The first is to use Bayesian analysis to estimate structural equation models (SEM). This helps manage small sample sizes \cite{lafifaRestructuringExpandingTechnology2012} and can produce a better data fit than frequentist approaches such as maximum likelihood estimation \cite{salarzadehjenatabadiTestingStudentsElearning2017}. The other direction is to use Bayesian networks \cite{kimUnderstandingFactorsInfluencing2025a}. These can be used to build decision-support systems and to predict technology acceptance within an organization \cite{kambleMachineLearningBased2021}. A key limitation of Bayesian networks is the inability to distinguish latent variables from their measurement items \cite{kambleMachineLearningBased2021}. However, this can be addressed by combining Bayesian networks with SEM \cite{guptaLinkingStructuralEquation2008}.

There are additional advantages to using Bayesian analysis compared to the frequentist approach. Furia et al.~\cite{furiaBayesianDataAnalysis2019} describe these in depth for empirical SE. First, the prior allows prior knowledge to be incorporated directly into the model. This prior knowledge can include domain expertise or findings from previous studies. As a result, researchers can incrementally build on existing research rather than starting from scratch, which is common in frequentist analysis. Additionally, researchers can test different priors to evaluate how different initial assumptions affect outcomes, leading to a more nuanced analysis. Second, Bayesian models can be updated with new evidence, allowing an existing model to be fine-tuned for a specific context and iteratively improved, even with limited data. Third, because the posterior distribution in Bayesian analysis represents probabilities, researchers can sample directly from it. For example, with a quantitative Likert-scale questionnaire, this approach enables predictions about outcomes based on specific responses and simulation of the effects of different choices. Overall, these benefits are exclusive to Bayesian analysis and offer a more flexible, nuanced approach to data analysis \cite{furiaBayesianDataAnalysis2019}.

UTAUT plays a crucial role in transitioning to Bayesian analysis. One of the key challenges with the Bayesian approach is the necessity to model the data domain \cite{furiaBayesianDataAnalysis2019}. Fortunately, UTAUT already provides a robust model to study technology acceptance. This allows us to base our assumptions about which constructs to measure and how they relate to one another on the strong theoretical foundation of UTAUT. The model can then evolve as we continue to study GenAI acceptance in SE.

\begin{figure*}[t]
\centering
\begin{tikzpicture}[
  font=\small,
  node distance=1mm and 10mm,
  c_utaut/.style={
    draw, rounded corners, align=center,
    minimum width=28mm, minimum height=6mm, inner sep=2mm,
    fill=white
  },
  c_utauttwo/.style={
    draw, rounded corners, align=center,
    minimum width=28mm, minimum height=6mm, inner sep=2mm,
    fill=gray!15
  },
  c_add/.style={
    draw, rounded corners, align=center,
    minimum width=28mm, minimum height=6mm, inner sep=2mm,
    fill=gray!30
  },
  out/.style={
    draw, rounded corners, align=center,
    minimum width=30mm, minimum height=8mm, inner sep=2mm,
    fill=white
    },
  box/.style={
    draw,
    rounded corners,
    align=center,
    minimum width=28mm,
    minimum height=6mm,
    inner sep=2mm
  },
  ut1/.style={-Latex, thick},   
  ut2/.style={-Latex, thick, dashed},   
  new/.style={-Latex, thick, dotted},           
]

\node[box] (bi) {Behavioral\\Intention};

\node[box, right=30mm of bi, yshift=-20mm] (use) {Actual\\Use};

\node[c_utaut, left=30mm of bi, yshift=20mm] (pe) {Performance\\Expectancy};
\node[c_utaut, below=of pe] (ee) {Effort\\Expectancy};
\node[c_utaut, below=of ee] (si) {Social\\Influence};
\node[c_utaut, below=of si] (fc) {Facilitating\\Conditions};
\node[c_utauttwo, below=of fc] (hm) {Hedonic\\Motivation};
\node[c_utauttwo, below=of hm] (hb) {Habit};
\node[c_add, below=of hb] (tc) {Technology\\Compatibility};
\node[c_add, below=of tc] (pi) {Personal\\Innovativeness};
\node[c_add, below=of pi] (ct) {Costs of\\Technology};
\node[c_add, below=of ct] (tr) {Trust};

\draw[ut1] (pe.east) -- (bi.west);
\draw[ut1] (ee.east) -- (bi.west);
\draw[ut1] (si.east) -- (bi.west);

\draw[ut2] (hm.east) -- (bi.west);
\draw[ut2] (hb.east) -- (bi.west);

\draw[ut2] (fc.east) -- (bi.west);
\draw[new] (tc.east) -- (bi.west);
\draw[new] (pi.east) -- (bi.west);
\draw[new] (ct.east) -- (bi.west);
\draw[new] (tr.east) -- (bi.west);

\draw[ut1] (bi.east) -- (use.west);
\draw[ut1] (fc.east) -- (use.west);

\draw[ut2] (hb.east) -- (use.west);

\node[anchor=north east, align=left, yshift=-15mm] at ($(use.south east)+(0mm,-8mm)$) {
\begin{tikzpicture}[baseline]
\node[] at (10mm,6mm) {Constructs};
\node[c_utaut, minimum width=10mm, minimum height=4mm, inner sep=1mm] (l1) {};
\node[right=2mm of l1] {\scriptsize UTAUT};

\node[c_utauttwo, minimum width=10mm, minimum height=4mm, inner sep=1mm, below=2mm of l1] (l2) {};
\node[right=2mm of l2] {\scriptsize UTAUT2};

\node[c_add, minimum width=10mm, minimum height=4mm, inner sep=1mm, below=2mm of l2] (l3) {};
\node[right=2mm of l3] {\scriptsize Additional};

\node[] at (40mm,6mm) {Relationships};
\draw[ut1] (20mm,-2mm) -- (30mm,-2mm);
\node[right=2mm] at (30mm,-2mm) {\scriptsize UTAUT};

\draw[ut2] (20mm,-8mm) -- (30mm,-8mm);
\node[right=2mm] at (30mm,-8mm) {\scriptsize UTAUT2};

\draw[new] (20mm,-14mm) -- (30mm,-14mm);
\node[right=2mm] at (30mm,-14mm) {\scriptsize Additional};
\end{tikzpicture}
};
\end{tikzpicture}
\caption{Constructs for GenAI acceptance in SE and their relationships to the constructs Behavioral Intention and Actual Use.}
\label{fig:utaut}
\end{figure*}
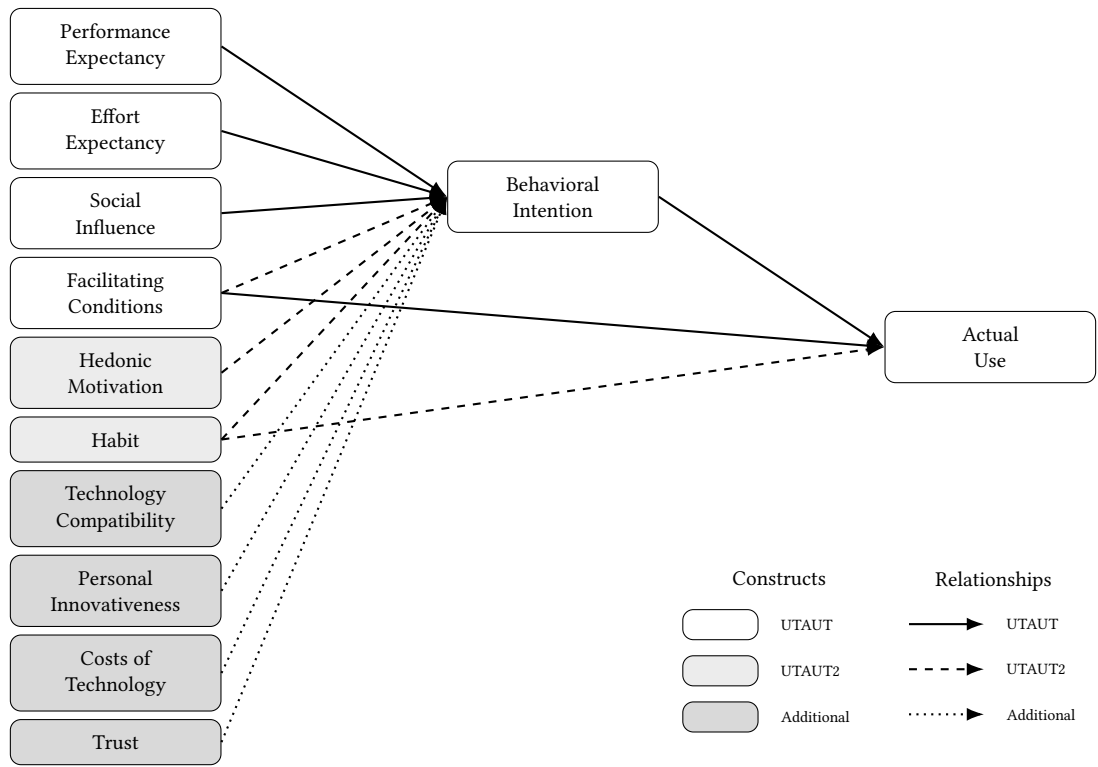

\section{Bayesian Analysis for Informed Decisions}
\label{sec:Example}

A Bayesian approach grounded in UTAUT can be useful in practice. By addressing the three research priorities outlined in this paper, we envision a process for making informed decisions about GenAI grounded in the individual perspectives of software engineers. In this section, we will exemplify this process using a fictitious scenario while explaining the roles of UTAUT and Bayesian analysis. 

\textbf{Scenario} A large-scale software development company is in the process of introducing an AI-based coding assistant. Due to low initial adoption of the tool, management aims to identify the most effective intervention to increase its actual use. To achieve this, the company has administered a Likert-scale questionnaire to measure the ten specific constructs outlined in Table~\ref{tab:constructs}.

\textbf{Bayesian analysis} A Bayesian network with the ten constructs is created, and we visualize the directed acyclic graph that the network is based on in Figure~\ref{fig:utaut}. The ten constructs are used to predict behavioral intention (i.e., \textit{acceptance}) and actual use of the coding assistant. We assume the same relationships as in UTAUT \cite{venkateshUserAcceptanceInformation2003}, UTAUT2 \cite{venkateshConsumerAcceptanceUse2012}, and additional literature \cite{parkContinuanceUseAI2025, choudhuriWhatGuidesOur2025b, russoNavigatingComplexityGenerative2024}. A prior based on the posterior distribution of a previous study (if such a study exists) at the company is used for the conditional probability distributions in the Bayesian network and updated with evidence from the new questionnaire. For the purpose of this scenario, we assume the analysis indicates that technology compatibility and facilitating conditions have a strong positive effect on behavioral intention and actual use, yet the actual questionnaire responses show low mean scores for these specific constructs.

\textbf{Follow-up} Focus groups with software engineers are held at the company to better understand the findings and identify appropriate interventions to improve technology compatibility and facilitating conditions. Two possible interventions are identified: improved IDE integration (technology compatibility) and additional training (facilitating conditions).

\textbf{Simulation} The company uses the Bayesian network to simulate the effect of the two interventions. This can be done by simulating counterfactuals \cite{mcelreathStatisticalRethinkingBayesian2020}, or in other words, \textit{what-if} scenarios. For instance, multiple simulations are run with different fixed values for technology compatibility and facilitating conditions, and a plot is created to summarize the results. We assume the simulations show that actual use is increased more by improved IDE integration than by additional training. While there are only minimal gains from introducing both interventions.

\textbf{Decision} The managers see that introducing both interventions yields minimal advantages compared to improving IDE integration alone. Thus, they decide to concentrate on improving the IDE integration. The company continues to monitor actual use over time and notes an increase following the intervention. Alternatively, the intervention may not have achieved the desired effect. In either case, the model can be updated with this new evidence and improved iteratively.

The purpose of this fictitious scenario is to provide one example of the increased options available to researchers and practitioners with a Bayesian approach. Bayesian analysis facilitates the exploration of \textit{what-if} scenarios. It is possible to test different priors and simulated scenarios to gain a more nuanced view before making decisions. The model can also be extended with additional evidence as new directions emerge, e.g., after a focus group. This can be used by software engineers to systematically compare alternative approaches or interventions and make evidence-based decisions related to GenAI in SE. To this end, UTAUT provides a solid theoretical foundation to build on. 

\section{Conclusion and Future Work}

GenAI is becoming an essential part of SE practice, making it increasingly important to understand why and how software engineers accept and use them. In this paper, we propose a research agenda to improve the relevance and applicability of technology acceptance models for GenAI acceptance in SE, using UTAUT as an example. Specifically, we advocate for the use of Bayesian analysis to study GenAI acceptance in SE. A Bayesian approach can offer greater flexibility and more nuanced insights than the commonly used frequentist approach. Additionally, UTAUT offers a strong theoretical foundation for incorporating prior knowledge into the model. The proposed approach aims to help software development managers make more informed decisions regarding GenAI adoption and software process improvement. It also helps researchers in designing empirical studies on technology acceptance.

To summarize, we identify three priorities for future work: (1) there is a need to identify the most meaningful constructs for studying GenAI acceptance in SE, (2) the operationalization of these constructs needs to be revisited to ensure it is meaningful in a modern SE context, and (3) Bayesian approaches should be explored for GenAI acceptance across various SE contexts. Lastly, the ideas presented in this paper may also be explored for other technologies beyond GenAI. 

\newpage

\section*{Acknowledgment}

This work has been supported by ELLIIT, a Strategic Area within IT and Mobile Communications, funded by the Swedish Government. The work has also been supported by a research grant for the GIST project (reference number 20220235) from the Knowledge Foundation in Sweden.

\bibliographystyle{ACM-Reference-Format}
\bibliography{references}

\end{document}